\documentclass[12pt,preprintnumbers]{revtex4-2}
\usepackage{amsfonts}
\usepackage{amsmath}
\usepackage{amssymb}
\usepackage{graphicx}
\usepackage{multirow}
\usepackage[colorlinks=true]{hyperref}
\usepackage{caption}
\usepackage{subcaption}
\newcommand{\al}{\alpha}
\newcommand{\non}{\nonumber}

\newcommand{\re}[1]{(\ref{#1})}

\newcommand{\veps}{\varepsilon}
\newcommand{\rar}{\rightarrow}

\begin{document}

\title{HCl, DCl and TCl diatomic molecules in their ground state: predicting
Born-Oppenheimer rovibrational spectra}

\date{\today}

\author{Horacio~Olivares-Pil\'on}
\email{horop@xanum.uam.mx}
\affiliation{Departamento de F\'isica, Universidad Aut\'onoma Metropolitana-Iztapalapa,
Apartado Postal 55-534, 09340 M\'exico, D.F., Mexico}

\author{Alexander~V.~Turbiner}
\email{turbiner@nucleares.unam.mx (Corresponding Author)}
\affiliation{Instituto de Ciencias Nucleares, Universidad Nacional
Aut\'onoma de M\'exico, Apartado Postal 70-543, 04510 M\'exico, D.F., Mexico}

\begin{abstract}
The analytic Born-Oppenheimer (B-O) potential curve for the ground state $X^1\Sigma^+$
of the molecule (H,D,T)Cl is constructed for the whole range
of internuclear distances $R \in [0,\infty)$ with an accuracy of 3-5 figures
in comparison with the RKR-style potential curve derived from available
experimental data on vibrational energies.
With an accuracy of 3-4 significant figures in the energies, it is predicted for HCl (DCl, TCl)
the 836 (1625, 2366) B-O rovibrational bound states with maximal
vibrational number $\nu_{max} = 20\, (29, 35)$ and maximal angular momentum
$L_{max} = 64\, (90, 109)$ including 24 (46, 63) weakly-bound states
(close to the dissociation limit) with energies $\lesssim 10^{-4}$\,Hartree.
Insufficiency of existing experimental data is indicated and a prediction
of the bulk of missing rovibrational states is made
for all HCl, DCl, TCl molecules.

\end{abstract}

\maketitle


\section{Introduction}
It is common knowledge that for a diatomic molecule the B-O
potential energy curve (saying differently, the electronic term, the electronic
energy curve) is defined as the eigenvalue of the electronic Hamiltonian,
when the nuclear masses are assumed to be infinite in the total Coulomb Hamiltonian
and the internuclear distance $R$ becomes a {\it classical}, non-dynamical variable.
Naturally, the eigenvalue (or, equivalently, the B-O potential energy curve)
depends on $R$ as a parameter and has no dependence on the nuclear masses.
Recently \cite{TO:2022}, it was proposed to consider the B-O potential energy curve as
the result of the screening of the Coulomb interaction of the two nuclear charges
$Z_A, Z_B$, taken as the probes,
\begin{equation}
\label{hetero-potential}
     V(R)\ =\ \frac{Z_A Z_B}{R}\,S(R)\ ,
\end{equation}
where $S(R)$ is the screening function, due to electronic media. At small distances $R$
the screening is negligible: the nuclei repel each other as bare Coulomb charges,
while at large $R$ the screening becomes an anti-screening: the van der Waals
attraction occurs instead of the Coulomb repulsion. Evidently, there always exists
an instance when the Coulomb repulsion is changed into the van der Waals attraction:
at this instance the nuclei do not interact and the potential energy curve has a minimum.
If this minimum is deep enough, strongly-bound states can occur. This manifests
the existence of the molecule. Needless to say that the potential energy curve
can be expanded around its minimum in a Taylor series: sometimes, the is called the Dunham expansion. We will call the coefficients of this expansion as the spectroscopic
(anharmonic) constants.

The screening function $S(R)$ is known theoretically in two limiting cases:
in the perturbation theory in small $R$ and in the form of the multipole expansion
(generated by induced moments) at large $R$, which is complimented by
exponentially-small terms for homonuclear molecules due to tunneling between the two identical Coulomb wells. In \cite{TO:2022}
a matching procedure for these three expansions into a single function was proposed,
which for the case of heteronuclear neutral molecules is realized in the form
of a two-point Pad\'e approximant = the ratio of two polynomials,
\begin{equation}
\label{SR}
     S(R)\ =\ P(N/N+5)\ \equiv \ \frac{P_N}{Q_{N+5}}\ ,
\end{equation}
where the integer parameter $N>1$ is chosen accordingly. This procedure can be applied
to any neutral diatomic molecule independently on the number of electrons and can be easily
extended to the diatomic molecular ions replacing $P(N/N+5)$ by $P(N/N+3)$.

This compact formula (\ref{SR}) was successfully applied to several neutral heteronuclear molecules -
HeH \cite{OT:2018}, LiH \cite{TO:2022}, ClF \cite{OT:2022}, H(D,T)F \cite{AG-OP:2022} -
to find the approximate B-O potential energy curves and then their associated B-O rovibrational spectra by solving the nuclear radial Schr\"odinger equation with the {\it standard} centrifugal potential $L(L+1)/R^2$ in addition to the vibrational B-O potential curve. Whenever possible this allows us to reproduce 3-5 significant figures of the experimental data on transition energies in the rovibrational spectra, cf.\cite{CH:2015}
as for the H(D,T)F molecules.
It is worth noting that in the case of homonuclear molecules, by adding to potential (\ref{hetero-potential}), (\ref{SR}) the extra exponentially-small terms at large $R$, see \cite{TO:2022}, complemented with the {\it standard} centrifugal potential $L(L+1)/R^2$,
this allows us to calculate the rovibrational energies with 3-5 significant figures in agreement with the experimental data for H${}_2^+$ and H$_2$ \cite{OT:2018}, He${}_2^+$ \cite{TO:2022}.

In general, the accuracy of the Born-Oppenheimer approximation is not well-understood \cite{Sutcliff}, see also \cite{Sutcliff-Woolley}.
The potential energy curve exists for the static case of infinitely-massive nuclei
{\it only}, when the internuclear distance is classical and can be varied ``manually".
If the masses of the nuclei are finite, the internuclear distance becomes a dynamical variable while its expectation value is an observable. By considering the nuclear kinetic energy in the total Coulomb Hamiltonian as a perturbation of the electronic Hamiltonian one can see that the deviation of the eigenvalues is of the order of the ratio of the masses of the electron and proton, $m_e/m_p \sim 5 \times 10^{-4}$, for the case of the ${\rm H}\,{}^{35} {\rm Cl}$ molecule.
Since the small parameter $m_e/m_{nucleus}$ appears in front of the second derivative of the total Hamiltonian the phenomenon of {\it bifurcation} occurs, which makes the perturbative analysis complicated. This was studied for the case of the exactly-solvable 3-body harmonic molecule \cite{TME:2020}.
It could be an indication that the mass corrections can change the 4th-5th significant figure
in the rovibrational energies.

It is assumed naturally that the first three-four-five significant figures of the rovibrational energies are correction-free: they are not influenced by non-adiabatic (mass corrections), also relativistic $(v/c)^2-$ and QED $\alpha^2-$ corrections (and any other corrections). This was checked for the case of the Helium-like and Lithium-like sequences in atomic physics \cite{Turbiner:2018},
the molecular ion H$^+_2$ \cite{OT:2018} and the H$_2$ molecule \cite{Komasa:2019} (and references therein). In the case of the (H,D,T)Cl molecules we continue to assume that
these correction-free significant figures in the rovibrational energies are given by the ones in the B-O rovibrational energies.

The goal of the present paper is to find the screening function $S(R)$ for
the ground state $X^1\Sigma^{+}$ of the hydrogenic halide ${\rm H}\,{}^{35} {\rm Cl}$,
and its two isotopologues ${\rm D}\,{}^{35} {\rm Cl}$, ${\rm T}\,{}^{35} {\rm Cl}$, and
to calculate the B-O rovibrational spectra for all three systems, respectively. The results will be compared with the experimental and theoretical/phenomenological data presented in the summary paper by Coxon-Hajigeorgiou \cite{CH:2015}, where the results of practically all previous studies of these halides are collected.

Throughout the paper the distances are given in atomic units (a.u.) while the energies are in Hartree.

\section{Potential energy curve}

Since the potential curve for diatomic molecules is not measurable experimentally,
there are two ways to construct it on scientific grounds:
(i) by solving the eigenvalue problem for the electronic Hamiltonian making the so-called
{\it ab initio} calculations and (ii) by solving the inverse problem of the quantum mechanics by taking the experimental transition (ro)vibrational energies as entry. The first way works well for one-, two-, three-electron diatomics but it leads to serious technical difficulties for more than three electron cases. The second way is based usually on the celebrated Rydberg-Klein-Rees (RKR) procedure or on its subsequent modifications, see \cite{LeRoy} and references therein. This procedure is ambiguous: an obtained potential is defined up to isospectral deformation. In order to remove this ambiguity, additional data like spectroscopical constants, transition amplitudes, scattering phases etc should be used.

In both ways the potential energy curve is (re)constructed in a point-wise manner, at a number of discrete points over $R$: this requires making an interpolation between points on the curve and an extrapolation to large and small internuclear distances. Additionally, there exists a phenomenological, extremely popular way to construct the potential energy curve near the equilibrium position by choosing an easily-treatable few-parametric potential which describes several low-lying (ro)vibrational energies or a few spectroscopical constants, see for review \cite{IK:2006,AB:2021} and references therein. Typically, these potentials do not describe correctly the behavior of the potential curves at both large and small internuclear distances, hence, the highly-excited rovibrational states close to dissociation limit, to a continuum spectra.

Hydrogen chloride HCl is made from two heavy nuclei: the proton $p$ and $cl$-nucleus,
and 18 electrons. At large distances between nuclei $R$ (the dissociation limit)
it is composed of two neutral atoms:
the Hydrogen atom H ($Z_{\rm H}=1$) and the Chlorine atom Cl ($Z_{\rm Cl}=17$).
The potential energy curve $E_d(R)$ in the lowest energy state of the molecular system
is related to the total energy $E_{total}$ as
\begin{equation}
    E_d(R)\ =\ E_{total}(R)\ -\ (E_{\rm H}+E_{\rm Cl})\ ,
\end{equation}
where $E_{\rm H}=-0.5$~a.u. and  $E_{\rm Cl}=-460.148$~a.u. \cite{CGD:1993}
are the ground state energies of the H- and Cl-atoms, respectively. In the united
atom limit $R \rar 0$, the nuclei become "glued" into a single nucleus and the HCl molecule
corresponds to the Argon atom Ar, where the ground state energy is
$E_{\rm Ar}=-527.540$~a.u. \cite{CGD:1993}.
The potential energy $E_d$ {\it vs.} $R$ defines the B-O potential energy curve, $V(R)=E_d$
with a normalization $V(R) \rar 0$ when $R \rar \infty$.

It is known that at small internuclear distances $R \rar 0$  the potential
energy curve admits the expansion
\begin{equation}
\label{EsmallR}
  E_d\ =\ \frac{17}{R}\ +\ \veps_0\ +\ 0 \cdot R\ +\ O(R^2)\ ,
\end{equation}
where the first term is the Coulomb repulsion potential $Z_{\rm H}Z_{\rm Cl}/R$
and $\veps_0$ is the reduced ground state energy of the Argon atom:
$\veps_0=E_{\rm Ar} +|E_{\rm H}+E_{\rm Cl} |=-66.892$~a.u.
Following Bingel~\cite{Bingel:1958} the linear term $\sim R$ is absent and
the higher order terms can contain powers of $\log R$, see also \cite{BBS:1966}.
The latter can be considered as an indication of the singular nature of the coefficients
in the expansion (\ref{EsmallR}) and, probably, of the zero radius of convergence
of the {\it amputated} function $\left(E_d - \frac{17}{R} \right)$ in the expansion at $R=0$.

Following Margenau-Pauling, see \cite{Pauling:1935} and references into it, and also for discussion \cite{MK:1971}, the behavior of the potential energy
at large internuclear distances $R \rightarrow \infty$,
\begin{equation}
\label{ElargeR}
  E_d\ =\ -\frac{C_6}{R^6}\ +\ \frac{C_8}{R^8}\ +\ \cdots\ ,
\end{equation}
corresponds to the multipole expansion of induced moments.
Note that the present authors are not familiar with the statement about the radius
of convergence of this expansion at $R=\infty$.
Here $C_6=23.41$\,a.u. is the van der Waals constant, based on the fit of the experimental
data, see \cite{Mitroy:2009}. Note that although this value differs from the Slater-Kirkwood
estimate $C_6=23.62$\,a.u., they are in agreement of $< 1\%$~, see for discussion
\cite{CH:2015}: this difference does not lead to qualitative consequences for the B-O rovibrational spectra, four significant figures in the rovibrational energies remain unchanged.
Matching the expansions (\ref{EsmallR}) and (\ref{ElargeR}) into a single function
leads naturally to the B-O potential energy curve in the form of a two-point Pad\'e approximant,
\begin{equation}
\label{Ed}
     E_d(R)\ =\ \frac{1}{R}\,{\rm Pade}[N /N + 5](R)_{n_0,n_{\infty}}\ \equiv \
     \frac{1}{R}\, \frac{P_N}{Q_{N+5}}\ ,
\end{equation}
where $P,Q$ are polynomials in $R$, here $n_0$ is the number of coefficients in the expansion
(\ref{EsmallR}) which we want to reproduce exactly, while $n_{\infty}$ is
the number of coefficients which are reproduced exactly at $R=\infty$ in the expansion
(\ref{ElargeR}). Following \cite{TO:2022} we will call the potential (\ref{Ed})
the Turbiner-Olivares (TO) potential.

As a first attempt to construct the B-O potential curve we choose $N=1$ in (\ref{Ed})
\begin{equation}
\label{fitP16}
    E^{(1)}_d(R)\ =\ \frac{1}{R}\,{\rm Pade}[1/6](R)_{2,1}\ =\
\end{equation}
\[
    \frac{1}{R}\ \frac{17 - a_1 R}
    {1 - \frac{(a_1+\veps_0)}{17}\, R + b_2 R^2\, +\, b_3 R^3 + b_4 R^4
     + b_5 R^5 + \frac{a_1}{C_6}\, R^{6}} \ ,
\]
where the five free parameters are fixed as follows
\[
  a_1 = 9.907\ ,\ b_2 = 19.493\ ,\ b_3 = -30.706\ ,\ b_4 = 17.952\ ,\ b_5 = -4.480 \ ,
\]
and $C_6=23.41$\,a.u.
This simple expression (\ref{fitP16}) allows us to reproduce two-three significant figures
in all available RKR-style data on the ground state potential curve collected in \cite{CH:2015}, see Table 9. However, in order to increase the accuracy and to reproduce four
(or more) significant figures with the five imposed conditions on the coefficients
in the expansions: $n_0=3, n_{\infty}=2$, higher order Pad\'e approximants
must be considered.

Concrete calculations were attempted to carry out with $N=3,4,5$. It is worth noting the following technical moment. Since $E_d$ is known at discrete points in $R$, two extra requirements were ``enforced" in order to find $S(R)$:
(i) the polynomial $Q(R)$ in (\ref{Ed}) has no roots at $R > 0$, hence, it is positive,
$Q(R) > 0$ at $R > 0$ and
(ii) the complex roots of $Q(R)$ (if exist) should be characterized either
by sufficiently large imaginary parts in the case of positive real part or to have
negative real parts in order to guarantee a sufficiently smooth monotonous behavior of
the potential energy curve at positive $R$. Implicitly, we required that $E_d$ has a single minimum at $R > 0$.
Eventually, we arrived at the fitting function (\ref{Ed}) with $N=5$ of the form
\begin{equation}
\label{fitP49}
    E^{(2)}_d(R)\ =\
    \frac{1}{R}\,\frac{17 +\, \sum_{i=1}^{4}\, a_i R^i-a_5 R^5}
    {1+\al_1 R + \al_2 R^2\, +\,\sum_{i=3}^{8}b_i R^i -\al_3 R^9+\al_4\, R^{10}} \ ,
\end{equation}
with the following constraints  imposed on the coefficients
\begin{eqnarray}
\label{par-1}
     \al_1 & =  & (a_1-\veps_0)/17\ ,\non \\
     \al_2 & =  & (\veps_0^2 + 17\, a_2  - a_1\veps_0)/17^2\ ,\non \\
     \al_3 & =  & a_4\, b_{8}/a_{5}\ ,\non \\
     \al_4 & =  & a_5/C_6\ ,
\end{eqnarray}
which guarantee that the three theoretically known coefficients in front of the
$R^{-1}$, $R^0$ and $R$ terms at small internuclear distances~\re{EsmallR} and
the two coefficients in front of $R^{-6}$ and $R^{-7}$ for large internuclear
distances~\re{ElargeR} are reproduced exactly. The remaining 11 parameters in
(\ref{fitP49}) are assumed to be free.
These parameters are fixed by making a fit of RKR-style experimental data on the ground state
potential energy curve for H(D,T)Cl, found in~\cite{CH:2015}, see also \cite{LeRoy}: their explicit values are
\begin{equation}
\begin{array}{lrrr}
 a_1 = &  327.321 \ ,    &\hspace{0.5cm} b_3 = & 33.7337\ ,\\
 a_2 = &  133.987 \ ,    & b_4 = &  -31.2621\ ,\\
 a_3 = & -275.248 \ ,    & b_5 = &  -7.72363\ ,\\
 a_4 = &  54.4855 \ ,    & b_6 = &   61.8757\ ,\\
 a_5 = &  3.611634\ ,  	 & b_7 = &  -44.2838\ ,\\
       &                 & b_8 = &   14.4997\ ,\\
\end{array}
\label{par-2}
\end{equation}
leading to an accuracy of the description in 4-5 significant figures. This implies that the spectroscopic
constants, equivalently, the coefficients in the Dunham expansion are reproduced with
a similar accuracy.
Alternately, three out of the 11 parameters in (\ref{par-2}) can be fixed by requiring
to reproduce 4-5 significant figures for the equilibrium distance, the
potential energy at equilibrium and zero vibrational energy
(the curvature of the potential well at the minimum).
It must be emphasized that in order to fix the 11 parameters (\ref{par-2}), in general,
it is enough to know RKR-style experimental data for the potential curve and/or the results of the {\it ab initio} calculations for 11 points in $R$ {\it only}.

\begin{figure}[h!]
\includegraphics[scale=2.0]{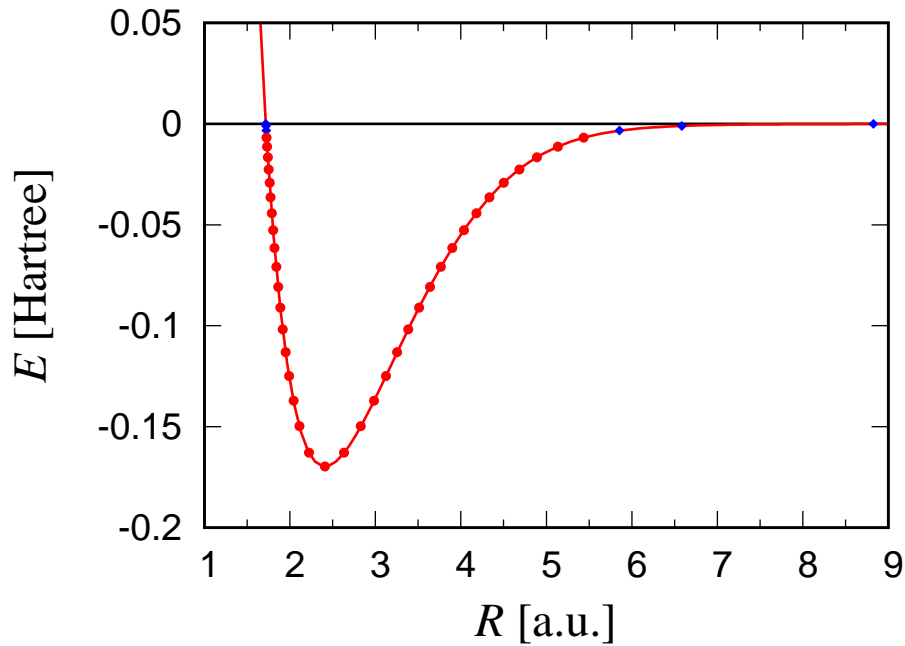}
\caption{\it Potential energy $E$ of the ground state $X^1\Sigma^+$ for the
   (H,D,T)Cl molecule {\it vs.} internuclear distance $R$:
   $(i)$ the Pad\'e approximant~\re{fitP49} (red solid line), $(ii)$  RKR-style
   data~\cite{CH:2015} (red bullets), $(iii)$ blue diamonds - predictions by \re{fitP49},
   $(iv)$ the energy $E=0$ at $R_0=1.71757$\,a.u. }
\label{FVpotClF}
\end{figure}

Table ~\ref{TcompV} presents the potential energy $E^{(2)}_d$ {\it vs.} $R$,
derived from the analytic expression~\re{fitP49} with parameters
(\ref{par-1})-(\ref{par-2}) for different internuclear distances $R$.
These are compared with all RKR-style experimental data and their phenomenological interpretation, see \cite{CH:2015}, Eq.(3) and Table 9:
as can be seen, except for a few points, not less than 4 decimal digits are reproduced.
Fit~\re{fitP49} is depicted in Fig.~\ref{FVpotClF} together with all RKR-style experimental
points, wherever they are available~\cite{CH:2015}. It must be emphasized
that such a quality of the description via ~\re{fitP49} does not require the
introduction of the terms $\sim (m_e/m_p)$ and $\sim (m_e/m_{cl})$
into the effective potential curve as was done in~\cite{CH:2015}, see Eq.(3).
Theoretically, the appearance of these terms contradicts the founding principles
of the Born-Oppenheimer approximation.

The minimum of the potential energy curve $E^{(2)}_d(R)$ is calculated
by taking the derivative of~\re{fitP49} and making it vanish, $dE^{(2)}_d(R)/dR=0$.
As a result $R_{min}=2.408542$\,a.u. and $E_{min}=-0.16960$\,Hartree, which is
in excellent agreement with the  values $R_{equilibrium}=2.408544$\,a.u.
and $E_{equilibrium}=-0.169641$\,Hartree~\cite{CH:2015}.

\begin{table}[h!]
\caption{Potential energy curve $E^{(2)}_d$ in Hartree for the ground state
$X^1\Sigma^+$ for the HCl molecule as a function of the internuclear distance $R$
in two regimes:
from small $R$ to equilibrium and from large $R$ to equilibrium.
2nd and 5th columns are RKR-style experimental data
from~\cite{CH:2015}. 3rd and 6th columns present data from~\re{fitP49}.}
\begin{center}
\begin{tabular}{c| cc| c| cc}
\hline\hline
$R_{min}$& $E_d^{R_{min}}$~\cite{CH:2015} & $E_d^{R_{min}}$~\re{fitP49}&
$R_{max}$& $E_d^{R_{max}}$~\cite{CH:2015} & $E_d^{R_{max}}$~\re{fitP49}\\
\hline
1.717716& -0.000082& -0.00011& 8.825021& -0.000082& -0.00007\\
1.719071& -0.001031& -0.00105& 6.585696& -0.001031& -0.00105\\
1.722427& -0.003364& -0.00338& 5.855864& -0.003364& -0.00341\\
1.727531& -0.006866& -0.00688& 5.437063& -0.006866& -0.00687\\
1.734145& -0.011318& -0.01132& 5.134002& -0.011318& -0.01128\\
1.742171& -0.016584& -0.01658& 4.890604& -0.016584& -0.01655\\
1.751549& -0.022571& -0.02256& 4.683618& -0.022571& -0.02256\\
1.762308& -0.029214& -0.02920& 4.500867& -0.029214& -0.02922\\
1.774500& -0.036461& -0.03645& 4.335109& -0.036461& -0.03649\\
1.788227& -0.044276& -0.04426& 4.181679& -0.044276& -0.04430\\
1.803636& -0.052627& -0.05261& 4.037339& -0.052627& -0.05264\\
1.820931& -0.061495& -0.06148& 3.899720& -0.061495& -0.06149\\
1.840391& -0.070860& -0.07085& 3.766980& -0.070860& -0.07084\\
1.862397& -0.080712& -0.08071& 3.637572& -0.080712& -0.08068\\
1.887477& -0.091041& -0.09105& 3.510072& -0.091041& -0.09101\\
1.916400& -0.101842& -0.10186& 3.383033& -0.101842& -0.10183\\
1.950315& -0.113112& -0.11313& 3.254758& -0.113112& -0.11311\\
1.991106& -0.124850& -0.12488& 3.122935& -0.124850& -0.12487\\
2.042201& -0.137055& -0.13708& 2.983789& -0.137055& -0.13709\\
2.111183& -0.149731& -0.14974& 2.829445& -0.149731& -0.14975\\
2.224536& -0.162880& -0.16286& 2.633188& -0.162880& -0.16287\\
2.408544& -0.169641& -0.16960&         &          &         \\
\hline\hline
\end{tabular}
\end{center}
\label{TcompV}
\end{table}

\section{B-O Rovibrational spectra}

In the standard Born-Oppenheimer approximation scheme, the rovibrational spectrum of HCl is found by solving the nuclear (radial) Schr\"odinger equation
\begin{equation}
\label{NucSc}
\left[-\frac{1}{2\mu_h}\frac{d^2}{dR^2}\ +\ \frac{L(L+1)}{2\mu_h R^2}\ +\ V(R)\right]\,\phi(R)\
=\  E_{\nu,L}\, \phi(R)\ ,
\end{equation}
searching for square-integrable eigenfunctions. Here
$\mu_h=m_{\rm Cl} m_{\rm H}/(m_{\rm Cl}+ m_{\rm H})$ is the reduced mass for the chlorine
and hydrogen nuclei, $L$ is the total angular momentum; $E_{\nu,L}$ is the rovibrational energy
of the state with vibrational and rotational quantum
numbers $\nu$ and $L$, respectively. The nuclear potential $V(R)=E_d$ is the electronic
energy curve given by~\re{fitP49}. Nowadays, equation~\re{NucSc} can be easily solved
numerically with any desired accuracy, in particular, by using the so-called Lagrange-Mesh
Method~\cite{DB:2015} (for the code written in MATHEMATICA-12, see \cite{OT:2022},
their extended ArXiv version). Nuclear masses are chosen as $m_{\rm H} =1836.15267$~\cite{MT:2018} for H and $m_{\rm Cl}=63727.3189$ for $^{35}$Cl~\cite{AWT:2003}, measured in terms of the mass of the electron.

The obtained B-O vibrational spectra $E_{\nu,0}$ of the bound states is presented in
Table~\ref{TvSClF}. In total, there are 21 vibrational states ($\nu=0,\dots, 20$):
this amount is in agreement with the experimental numbers presented in~\cite{CH:2015}.
Compared with the 21 experimental vibrational energies reported in~\cite{CH:2015},
there is a coincidence in 4 decimal digits (d.d.). Note that the energy $E_{20,0}$
of the bound state (20,0) is close to the dissociation threshold because it is smaller than $10^{-4}$\,Hartree. Thus, it is a weakly-bound state. Its existence is certainly questionable depending upon the values of mass, relativistic and QED corrections.

The B-O rotational spectra $E_{0,L}$ of the bound states is presented in
Table~\ref{rotSHCl}. It coincides systematically in 4 decimal digits with experimental data \cite{CH:2015} (after rounding) for $L=0,1,2,\ldots, 50$. Let is note that for larger angular momenta $L=51, \ldots , 64$ the experimental data are absent and our results can be considered as a prediction.

In general, the B-O rovibrational energy spectra $E_{\nu,L}$ of the bound states reproduce systematically the 4 decimal digits of the experimental data wherever they are available, see Table \ref{TLLn0HCLb}, the values $L=10, 20, 30, 40, 50$ are presented as the example.

\begin{table}[h!]
\caption{B-O vibrational energies $E_{\nu,0}$ in Hartree for the ground state $X^1\Sigma^+$ of the HCl molecule. Vibrational energies from~\cite{CH:2015} are presented in the 3rd column.}
\begin{center}
\scalebox{1}{
\begin{tabular}{r  cc}
\hline\hline
$\nu$&$E_{\nu,0}$ &\cite{CH:2015} \\
\hline
0 & -0.16285& -0.162880\\
1 & -0.14972& -0.149731\\
2 & -0.13705& -0.137055\\
3 & -0.12485& -0.124850\\
4 & -0.11311& -0.113112\\
5 & -0.10183& -0.101842\\
6 & -0.09102& -0.091041\\
7 & -0.08068& -0.080712\\
8 & -0.07083& -0.070860\\
9 & -0.06147& -0.061495\\
10& -0.05260& -0.052627\\
11& -0.04425& -0.044276\\
12& -0.03644& -0.036461\\
13& -0.02919& -0.029214\\
14& -0.02255& -0.022571\\
15& -0.01655& -0.016584\\
16& -0.01128& -0.011318\\
17& -0.00684& -0.006866\\
18& -0.00336& -0.003364\\
19& -0.00104& -0.001031\\
20& -0.00007& -0.000082\\
\hline\hline
\end{tabular}}
\end{center}
\label{TvSClF}
\end{table}

\begin{table}[h!]
\caption{B-O rotational energies $E_{0,L}$ in Hartree for the ground state $X^1\Sigma^+$ of the HCl molecule for $L=0,\cdots,64$ compared with experimental data, see~\cite{CH:2015}.}
\begin{center}
\scalebox{1}{
\begin{tabular}{ rcc | lcc | lcc }
\hline\hline
$L$&$E_{0,L}$ &\cite{CH:2015}&$L$&$E_{0,L}$ &\cite{CH:2015}&$L$&$E_{0,L}$ &\cite{CH:2015} \\
\hline
0 &  -0.16285&  -0.162880& 22&  -0.13938&  -0.139417& 44&  -0.07755&  -0.077592\\
1 &  -0.16275&  -0.162785& 23&  -0.13730&  -0.137343& 45&  -0.07407&  -0.074115\\
2 &  -0.16256&  -0.162595& 24&  -0.13515&  -0.135189& 46&  -0.07055&  -0.070594\\
3 &  -0.16228&  -0.162310& 25&  -0.13292&  -0.132957& 47&  -0.06699&  -0.067032\\
4 &  -0.16190&  -0.161930& 26&  -0.13061&  -0.130647& 48&  -0.06338&  -0.063430\\
5 &  -0.16142&  -0.161456& 27&  -0.12822&  -0.128262& 49&  -0.05975&  -0.059791\\
6 &  -0.16086&  -0.160887& 28&  -0.12576&  -0.125802& 50&  -0.05607&  -0.056117\\
7 &  -0.16019&  -0.160224& 29&  -0.12323&  -0.123269& 51&  -0.05236&	       \\
8 &  -0.15944&  -0.159468& 30&  -0.12062&  -0.120664& 52&  -0.04863&	       \\
9 &  -0.15859&  -0.158619& 31&  -0.11795&  -0.117989& 53&  -0.04486&	       \\
10&  -0.15764&  -0.157677& 32&  -0.11520&  -0.115246& 54&  -0.04107&	       \\
11&  -0.15661&  -0.156643& 33&  -0.11239&  -0.112435& 55&  -0.03725&	       \\
12&  -0.15548&  -0.155518& 34&  -0.10951&  -0.109560& 56&  -0.03341&	       \\
13&  -0.15427&  -0.154302& 35&  -0.10951&  -0.106621& 57&  -0.02955&	       \\
14&  -0.15296&  -0.152996& 36&  -0.10357&  -0.103619& 58&  -0.02567&	       \\
15&  -0.15157&  -0.151601& 37&  -0.10051&  -0.100558& 59&  -0.02178&	       \\
16&  -0.15008&  -0.150118& 38&  -0.09739&  -0.097438& 60&  -0.01787&	       \\
17&  -0.14851&  -0.148547& 39&  -0.09422&  -0.094261& 61&  -0.01395&	       \\
18&  -0.14685&  -0.146890& 40&  -0.09098&  -0.091030& 62&  -0.01003&	       \\
19&  -0.14511&  -0.145148& 41&  -0.08770&  -0.087745& 63&  -0.00610&	       \\
20&  -0.14328&  -0.143320& 42&  -0.08436&  -0.084410& 64&  -0.00216&	       \\
21&  -0.14137&  -0.141410& 43&  -0.08098&  -0.081025&	&	   &	       \\
\hline\hline
\end{tabular}}
\end{center}
\label{rotSHCl}
\end{table}

\begin{table}[h!]
\caption{B-O ro-vibrational energies $E_{\nu,L}$ in Hartree for the ground state $X^1\Sigma^+$ of the HCl molecule. Values for $L=10, 20, 30, 40$ and $50$ are presented.  Experimental energies from ~\cite{CH:2015} are given in the second column.}
\begin{center}
\scalebox{0.8}{
\begin{tabular}{r|cc|cc|cc|cc|cc}
\hline\hline
$\nu$& $E_{\nu,10}$&\cite{CH:2015}&  $E_{\nu,20}$ &\cite{CH:2015} & $E_{\nu,30}$ &\cite{CH:2015} & $E_{\nu,40}$ &\cite{CH:2015} & $E_{\nu,50}$ &\cite{CH:2015} \\
\hline
0 & -0.15764& -0.157677& -0.14328& -0.143320& -0.12062& -0.120664& -0.09098& -0.091030& -0.05607& -0.056117\\
1 & -0.14466& -0.144679& -0.13073& -0.130747& -0.10875& -0.108779& -0.08004& -0.080083& -0.04630& -0.046354\\
2 & -0.13215& -0.132154& -0.11863& -0.118644& -0.09733& -0.097360& -0.06956& -0.069602& -0.03700& -0.037066\\
3 & -0.12009& -0.120099& -0.10699& -0.107009& -0.08638& -0.086407& -0.05954& -0.059592& -0.02819& -0.028267\\
4 & -0.10850& -0.108511& -0.09582& -0.095840& -0.07589& -0.075923& -0.05000& -0.050058& -0.01990& -0.019972\\
5 & -0.09737& -0.097391& -0.08511& -0.085140& -0.06587& -0.065912& -0.04095& -0.041010& -0.01214&          \\
6 & -0.08672& -0.086740& -0.07488& -0.074911& -0.05634& -0.056380& -0.03241& -0.032465& -0.00494&          \\
7 & -0.07653& -0.076562& -0.06512& -0.065161& -0.04729& -0.047339& -0.02439& -0.024443&         &          \\
8 & -0.06683& -0.066863& -0.05586& -0.055897& -0.03876& -0.038804& -0.01693& -0.016973&         &          \\
9 & -0.05762& -0.057654& -0.04710& -0.047134& -0.03076& -0.030795& -0.01005& -0.010097&         &          \\
10& -0.04892& -0.048949& -0.03886& -0.038890& -0.02331& -0.023342& -0.00383& -0.003873&         &          \\
11& -0.04074& -0.040764& -0.03116& -0.031188& -0.01645& -0.016484&         &  0.001612&         &          \\
12& -0.03310& -0.033126& -0.02404& -0.024061& -0.01023& -0.010274&         &  0.006208&         &          \\
13& -0.02604& -0.026066& -0.01752& -0.017553& -0.00474& -0.004789&         &          &         &          \\
14& -0.01960& -0.019626& -0.01168& -0.011721& -0.00009& -0.000146&         &          &         &          \\
15& -0.01383& -0.013864& -0.00660& -0.006648&         &  0.003434&         &          &         &          \\
16& -0.00882& -0.008855& -0.00242& -0.002457&         &          &         &          &         &          \\
17& -0.00468& -0.004709&         &          &         &          &         &          &         &          \\
18& -0.00160&          &         &          &         &          &         &          &         &          \\
\hline\hline
\end{tabular}}
\end{center}
\label{TLLn0HCLb}
\end{table}

\begin{figure}[h!]
\includegraphics[scale=2.4]{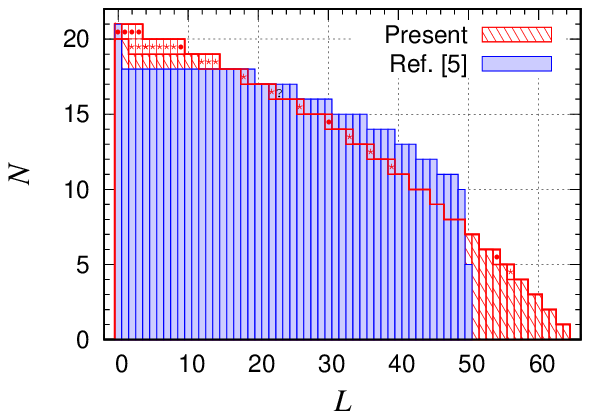}
\caption{\it Rovibrational spectra for the ground state $X^1\Sigma^+$ of hydrogen
chloride HCl - the histogram - emerged in the potential ~\re{fitP49} with standard centrifugal potential.
The 7 states with energy $< 10^{-4}$\,Hartree marked by bullets and
the 17 states whose energies are of order $\sim 10^{-4}$\,Hartree marked by stars.
The rovibrational states (shadowed in blue) are the results from \cite{CH:2015}.
Those states which are above the red, continuous, piece-wise line are from
the phenomenological analysis performed in \cite{CH:2015}: they have positive
energies.
The state $(\nu=17,L=23)$ with the energy $E_{17,23}$, labeled by "?" ,
indicated as bound in~\cite{CH:2015}: it is not confirmed
in the present calculation. All predicted B-O states with $L=51-64$ are absent
in the experiment, see \cite{CH:2015}.}
\label{rvsHCl}
\end{figure}

The complete rovibrational spectra is depicted in the histogram in
Fig.~\ref{rvsHCl}. In total, there are 836 B-O rovibrational states with
$\nu_{max} = 20$ and  $L_{max} = 64$. In general, there are 7 states with
energy $\lesssim 10^{-4}$\,Hartree and additionally 17 states whose energies
are of order $\sim 10^{-4}$\,Hartree - these are marked by dots and stars,
respectively, in the histogram. States above the red, continuous,
but piece-wise line are of positive energy, thus, they are quasi-bound states.
These are present in the phenomenological analysis~\cite{CH:2015} as well as
in experimental data which are beyond {\it our} B-O analysis. Note that
these quasi-bound states in \cite{CH:2015} might be a consequence
of the modification of the nuclear Schr\"odinger equation (\ref{NucSc})
proposed in \cite{CH:2015}, see Eqs.(1)-(3), where the potential curve
was not studied at $R < R_0$, see Fig.\ref{FVpotClF}.
The energy $E_{17,23}$, labeled by "?" in the histogram, is indicated
in~\cite{CH:2015} as the one of the bound state: it is not observed
in the present B-O calculation.
It must be emphasized that 57 B-O rovibrational states predicted
for $L=50 - 64$ are absent in experimental data as well as 26 highly-excited B-O rovibrational states for $L=1 - 14$. It might be an indication to insufficiency
of experimental data.

\begin{figure}[h!]
\includegraphics[scale=2.4]{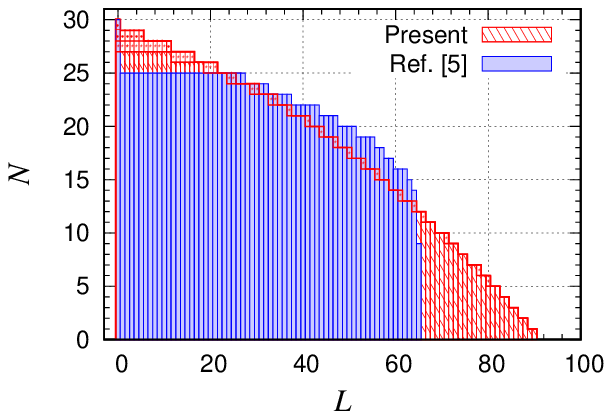}
\caption{\it Rovibrational spectra for the ground state $X^1\Sigma^+$ of deutron
chloride DCl - the histogram - emerged for the potential ~\re{fitP49} with standard centrifugal potential.
The 8 states (including 2 for $L=0$) with energy $< 10^{-4}$\,Hartree
marked by bullets and the 38 states (including 1 for $L=0$) whose energies
are of order $\sim 10^{-4}$\,Hartree marked by stars.
The rovibrational states (shadowed in blue) are the experimental (and phenomenological)
results from \cite{CH:2015}.
Those states which are above the red, continuous, piece-wise line are from
the phenomenological analysis \cite{CH:2015}, they have of positive
energies. All 157 states with $L=66-90$ are absent in the experimental
(and phenomenological) data \cite{CH:2015} as well as 53 states with $L=1-21$.}
\label{rvsDCl}
\end{figure}

\begin{figure}[h!]
\includegraphics[scale=2.4]{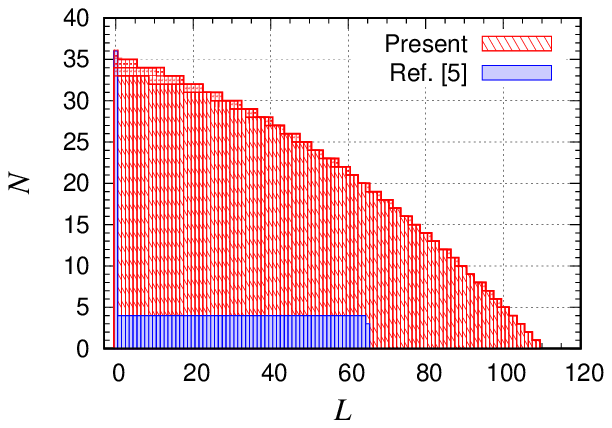}
\caption{\it Rovibrational spectra for the ground state $X^1\Sigma^+$ of triton
chloride TCl - the histogram - emerged for the potential ~\re{fitP49} with standard centrifugal potential.
The 15 states (including two with $L=0$) with energy $< 10^{-4}$\,Hartree marked
by bullets and the 48 states (including one with $L=0$) whose energies are of order $\sim 10^{-4}$\,Hartree marked by stars. The rovibrational states (shadowed in blue) are the experimental (and phenomenological) results from \cite{CH:2015}.
Almost all 2071 B-O rovibrational states with $L=1-109$ are absent in experimental
(and phenomenological) data~\cite{CH:2015}.}
\label{rvsTCl}
\end{figure}

The same B-O potential energy curve ~\re{fitP49} can be used to calculate the
B-O rovibrational spectra of the HCl isotopologues: the deutron chloride DCl and
the triton chloride TCl. This implies an appropriate change of the reduced mass
in the nuclear Schr\"odinger equation (\ref{NucSc}),
\[
  \mu_h \ \rar \ \mu_d= \frac{m_{\rm Cl}\ m_{\rm D}}{(m_{\rm Cl}+ m_{\rm D})} \ ,
\]
or
\[
  \mu_h \ \rar \ \mu_t= \frac{m_{\rm Cl} m_{\rm T}}{(m_{\rm Cl}+ m_{\rm T})} \ ,
\]
respectively. Here $m_{\rm D,T}$ are the masses of deutron $m_{\rm D}=\ 3670.482\,97$\,$m_e$
and triton $m_{\rm T}=\ 5496.921\,54$\,$m_e$, respectively, see~\cite{MT:2018}.

In Tables V and VI a comparison of the vibrational states for DCl and TCl is presented:
in general, there is a coincidence of 4 d.d. in the energies with the experimental data.
The highest vibrational energies $E_{29,0}$ for DCl and $E_{35,0}$ for TCl, respectively,
are smaller than $10^{-4}$\,Hartree, hence, they are close to the dissociation threshold. They should be excluded. Furthermore the vibrational energies $E_{28,0}$ for DCl and $E_{34,0}$ for TCl are of the order of $10^{-4}$\,Hartree: their existence is questionable,
the corrections could be of the same order of magnitude. Note that the maximal number
of vibrational states is increased from 19 for HCl to 28 for DCl and 34 for TCl,
respectively, see Table VII. In a similar way the maximal number of rotational states
is increased from 64 for HCl to 90 for DCl and 109 for TCl, respectively, see Table VII.

\begin{table}[h!]
\caption{Vibrational energies $E_{\nu,0}$ in Hartree for the ground state $X^1\Sigma^+$ of
  the DCl molecule. Experimental data for the energies of the vibrational
  states~\cite{CH:2015} are presented in 3rd and 5th columns. Energies $E_{28,0}$ and
  $E_{29,0}$ are beyond of our accuracy.}
\begin{center}
\scalebox{1}{
\begin{tabular}{l  lc | l  lc}
\hline\hline
$\nu$&\,\,\,\,\,\,\,$E_{\nu,0}$ &\cite{CH:2015}&
$\nu$&\,\,\,\,\,\,\,$E_{\nu,0}$ &\cite{CH:2015}\\
\hline
0 & -0.164749&  -0.164767& 15& -0.047403&  -0.047408\\
1 & -0.155238&  -0.155240& 16& -0.041596&  -0.041599\\
2 & -0.145966&  -0.145957& 17& -0.036069&  -0.036072\\
3 & -0.136932&  -0.136917& 18& -0.030833&  -0.030837\\
4 & -0.128136&  -0.128119& 19& -0.025901&  -0.025908\\
5 & -0.119578&  -0.119562& 20& -0.021289&  -0.021301\\
6 & -0.111258&  -0.111246& 21& -0.017018&  -0.017038\\
7 & -0.103178&  -0.103170& 22& -0.013116&  -0.013143\\
8 & -0.095338&  -0.095335& 23& -0.009615&  -0.009646\\
9 & -0.087740&  -0.087742& 24& -0.006562&  -0.006588\\
10& -0.080387&  -0.080393& 25& -0.004010&  -0.004024\\
11& -0.073281&  -0.073289& 26& -0.002033&  -0.002033\\
12& -0.066425&  -0.066434& 27& -0.000712&  -0.000718\\
13& -0.059824&  -0.059832& 28& -0.000095&  -0.000122\\
14 & -0.053481 & -0.053488 & 29 & -9.E(-9) & -2.E(-6) \\
\hline\hline
\end{tabular}}
\end{center}
\label{TvSDCl}
\end{table}

\begin{table}[h!]
\caption{Vibrational energies $E_{\nu,0}$ in Hartree for the ground state $X^1\Sigma^+$
 of TCl molecule. Experimental  and theoretical data for energies of the vibrational
 states~\cite{CH:2015} are presented in the 3rd and 5th columns. Values $E_{34,0}$ and $E_{35,0}$
 are beyond the present accuracy.}
\begin{center}
\scalebox{1}{
\begin{tabular}{l  lc | l  lc}
\hline\hline
$\nu$&\,\,\,\,\,\,\,$E_{\nu,0}$ &\cite{CH:2015}&
$\nu$&\,\,\,\,\,\,\,$E_{\nu,0}$ &\cite{CH:2015}\\
\hline
0 & -0.165579& -0.165593&18& -0.048473& -0.048472\\
1 & -0.157668& -0.157668&19& -0.043603& -0.043601\\
2 & -0.149921& -0.149910&20& -0.038924& -0.038921\\
3 & -0.142338& -0.142320&21& -0.034440& -0.034437\\
4 & -0.134918& -0.134896&22& -0.030157& -0.030156\\
5 & -0.127661& -0.127638&23& -0.026084& -0.026087\\
6 & -0.120568& -0.120545&24& -0.022231& -0.022238\\
7 & -0.113638& -0.113617&25& -0.018608& -0.018622\\
8 & -0.106871& -0.106855&26& -0.015229& -0.015250\\
9 & -0.100270& -0.100257&27& -0.012114& -0.012140\\
10& -0.093833& -0.093825&28& -0.009281& -0.009310\\
11& -0.087563& -0.087559&29& -0.006758& -0.006785\\
12& -0.081461& -0.081460&30& -0.004576& -0.004595\\
13& -0.075528& -0.075529&31& -0.002775& -0.002782\\
14& -0.069766& -0.069768&32& -0.001400& -0.001403\\
15& -0.064177& -0.064179&33& -0.000498& -0.000515\\
16& -0.058763& -0.058765&34& -0.000077& -0.000108\\
17 & -0.053527 & -0.053528 & 35 & -4.E(-7) & -6.E(-6)\\
\hline\hline
\end{tabular}}
\end{center}
\label{TvSTCl}
\end{table}

\begin{table}[h!]
\caption{$\nu_{max}$ and $L_{max}$ for the systems HCl, DCl and TCl
         obtained in the present work. The numbers in brackets
         (20), (29) and (36) indicate the total number of vibrational
         states, which includes the states with energy $< 10^{-4}$\,Hartree.
         Reduced mass of nuclei $\mu$ indicated.}
\begin{center}
\scalebox{0.9}{
\begin{tabular}{r | lcrc}
\hline\hline
      &\ $\nu_{max}$\ &\ $L_{max}$\ &\ Total\  &\ $\mu $\\
\hline
HCl \ &\ 19 (20)\ &\  64\ &  836 \ &\ 1784.7299 \\
DCl \ &\ 28 (29)\ &\  90\ & 1625 \ &\ 3470.5885 \\
TCl \ &\ 34 (35)\ &\ 109\ & 2366 \ &\ 5060.4249 \\
\hline\hline
\end{tabular}}
\end{center}
\label{XCl}
\end{table}

For DCl the complete rovibrational spectra is depicted in the histogram in
Fig.~\ref{rvsDCl}. In total, there are 1625 B-O rovibrational states with
$\nu_{max} = 29$ and  $L_{max} = 90$. In general, there are 8 states with
energy $\lesssim 10^{-4}$\,Hartree and additionally 38 states whose energies
are of order $\sim 10^{-4}$\,Hartree - in the histogram, they are marked
by bullets and stars, respectively. States above the red, continuous, piece-wise line
have positive energy, thus, they are quasi-bound states. These are present in
the phenomenological analysis~\cite{CH:2015} as well as in the experimental data but they
are not confirmed in the present B-O analysis.
It must be emphasized that 160 B-O rovibrational states predicted for $L=65 - 90$
are absent in the experimental data as well as 53 highly excited B-O rovibrational states
for $L=1 - 21$. We do not have any explanation of it. This might be a reflection of
insufficiency of the experimental data.

For TCl the complete rovibrational spectra is depicted in the histogram in
Fig.~\ref{rvsTCl}. In total, there are 2366 B-O rovibrational states with
$\nu_{max} = 36$ and  $L_{max} = 109$. In general, there are 15 states with
energy $\lesssim 10^{-4}$\,Hartree and additionally 48 states whose energies
are of order $\sim 10^{-4}$\,Hartree - they are marked by bullets and stars,
respectively, in the histogram. There are 3 vibrational states with
energies $\lesssim 10^{-4}$\,Hartree.
It must be emphasized that 2071 B-O rovibrational states are
predicted for $L=1 - 109$ - almost all of them are absent from the experimental
(and phenomenological) data.

{

\section{Discussion}

By using the physics arguments, which define the asymptotics \cite{TO:2022}, with addition of the experimental, RKR-style data for the potential energy curve mostly in the vicinity of the equilibrium configuration the Born-Oppenheimer potential energy curve for the ground state $X^1\Sigma^+$ of the (H,D,T)Cl molecules was constructed in the form of a ratio of two polynomials in the whole range of internuclear distances. By making the comparison with the RKR-style experimental (and phenomenological) data, where they are available,
one can see that the resulting B-O potential energy curve is in agreement within 3-5 significant figures in the whole domain in $R \in [0, \infty)$. We assume it is accurate in 4 significant digits.
This potential curve, when used as a potential in the nuclear radial Schr\"odinger equation
with standard centrifugal potential additionally, allows us
to calculate the B-O (ro)vibrational spectra for all three molecules HCl, DCl, TCl while
taking into account the dependence on the reduced mass of the two nuclei ONLY. By comparing
the B-O rovibrational energies with the available experimental (and phenomenolocal)
data one can see a general agreement in energies in 4-5 significant figures (or better).
Note that instead of presenting lengthy Tables with rovibrational energies
we provide a simple MATHEMATICA-12 code, see \cite{OT:2022}, ArXiv version: 2202.10666v2.
In particular, there is an excellent agreement in 4-5 significant figures
with the description of the vibrational energies of all three molecules HCl, DCl, TCl.
All this defines the domain of applicability of
the Born-Oppenheimer approximation, which is the domain free of corrections of any kind.

\section{Conclusions}

The situation with vibrational states $(L=0)$ looks satisfactory for all three molecules H(D,T)Cl:
vibrational energies are in complete agreement with B-O predictions in 4-5 s.d. However,
for low angular momenta $L=1 - 15 (HCl), 20 (DCl)$ there are many missing highly-excited
vibrational states predicted in B-O approximation, even though the low-excited vibrational states
are in agreement with experimental data in 3-5 s.d. Situation is opposite for larger angular
momenta $L > 15, 20$: there are highly-excited vibrational states seen experimentally, which B-O approximation does not predict although the low-excited vibrational states
are in agreement with experimental data.

The obtained results in B-O approximation seemingly indicate that a lot of experimental
information is missing:
83 rovibrational states for HCl, 212 rovibrational states for DCl, 2071 rovibrational
states for TCl. Our B-O approximation results can be considered as a prediction of the existence
of those states, many of which are highly-excited rovibrational bound states, in particular,
with large angular momenta, which is hardly accessible experimentally. It should be emphasized
that the experimental situation about the rovibrational spectra of the TCl molecule is extremely
suspicious: from physics point of view the measured rovibrational spectra can not be like one showed on
Fig.\ref{rvsTCl} even for small angular momenta.

Recently, a similar B-O analysis was conducted for the HF, DF, TF molecules,
see \cite{AG-OP:2022}, with analogous conclusions: a lot of experimental information
is missing especially for TF molecule. Predictions about missing rovibrational states were made there.

It is in our plans to make the Born-Oppenheimer analysis for the ground X${}^1 \Sigma$ state of the two remaining hydrogenic halids H(D,T)Br and H(D,T)I.

In upcoming work we are re-analyzing the $H(H,D,T)^+$ ions in B-O approximation in our approach see \cite{OT:2018} and in full geometry, when all masses are finite, by using the Lagrange-mesh method.
For the first two molecules $H_2^+$ and $HD^+$ the rovibrational spectra in both considerations are in complete agreement with existing calculations and, in particular, the rovibrational energies are in agreement within 4 s.d. The situation changes dramatically for $HT^+$. If our calculated vibrational
spectra $(L=0)$ and rovibrational ones at $L=1-5$ and $\nu=0-5$ are in agreement with Adamovicz et al and Korobov et al within 12-13 s.d., we predict about 400 more rovibrational states for larger $L>5$ and for vibrational excited states for $L>0$. It will be published elsewhere.

}

\section*{Acknowledgements}

The research is partially supported by CONACyT grant A1-S-17364 and DGAPA grant
IN113022 (Mexico). A.V.T. thanks PASPA-UNAM for a partial support during the sabbatical stay at University of Miami (2021-2022), where this work was mostly carried out. { With a great sadness we had learned that Prof. Brian Sutcliffe, with whom we had extremely fruitful mail correspondence, passed away on Dec.26, 2022.}

\end{document}